\documentstyle[12pt]{article}

\begin{document}
\def\bib#1{[{\ref{#1}}]}
\def\at{\tilde{a}}

\begin{titlepage}
         \title{ Maximal Acceleration Corrections  \\
                 to the Lamb Shift of One-Electron Atoms
\footnote{Research supported by MURST fund 40\% and 60\%, DPR 382/80}}

\author{G. Lambiase$^{a}$\thanks{{\it{E-mail}}: lambiase@vaxsa.csied.unisa.it}
,~~~G.Papini$^{b}$\thanks{{\it{E-mail}}: papini@cas.uregina.ca}~~~and
~~~G. Scarpetta$^{a,c}$ \\
{\em $^a$Dipartimento di Scienze Fisiche ``E.R. Caianiello''}\\ 
{\em  Universit\`a di Salerno, 84081 Baronissi (SA), Italy}\\
{\em  $^a$Istituto Nazionale di Fisica Nucleare, Sezione di Napoli}\\
{\em $^b$Department of Physics, University of Regina,} \\
{\em Regina, Sask. S4S 0A2, Canada}\\
{\em $^c$International Institute for Advanced Scientific Studies} \\ 
{\em Vietri sul Mare (SA), Italy}}
              \date{\empty}
              \maketitle

              \begin{abstract}
The maximal acceleration corrections to the Lamb shift of one--electron atoms 
are calculated starting from the Dirac equation and splitting the spinor into
large and small components. The results depend on $Z^8$ and a
cut-off $\Lambda$. Sizeable values are obtained even at $Z=1$ for 
$\Lambda\sim a_0/2$, where $a_0$ is the Bohr radius. These values are 
compatible with theoretical and experimental results.
	      \end{abstract}

\thispagestyle{empty}
\vspace{20. mm}
              \vfill
	      \end{titlepage}
\section{\bf Introduction}
\setcounter{equation}{0}

The notion of a limiting value to the proper acceleration of a particle, 
advanced on classical and quantum grounds by several authors \cite{ANI}, \cite{TUT}, \cite{BRA}, 
has different
aspects and formulations. The importance of its existence, if proven, cannot
be underestimated. It would by itself rid quantum field theory of unpleasant 
divergencies, and make several important theoretical procedures finite.

The idea finds a particularly interesting formulation in the original
attempts by Caianiello \cite{ANI} to incorporate quantum 
position-momentum uncertainty relations into the geometry of space-time,
and to interpret quantization geometrically as a consequence of curvature in 
phase space.

In this model a particle of mass $m$
accelerating along its worldline behaves dynamically as if subject to an
effective gravitational field
\begin{equation}
\tilde g_{\mu\nu}=g_{\mu\nu}\left(1-\frac{|\ddot{x}|^2}{A^2}
\right)\,{,}
\end{equation}
where $|\ddot{x}|^2 = |g_{\mu\nu} \ddot{x}^\mu \ddot{x}^\nu|$ is the square length of the 
relativistic acceleration four--vector,
 $A=mc^3/\hbar$ is the maximal proper acceleration of the particle and 
$g_{\mu\nu}$ represents a background gravitational field. 

Eq. (1.1), the arrival point of an embedding procedure  of an
eight--di\-men\-sional space--time tangent bundle TM in four--dimensional space--time \cite{GAS},
has several important implications for relativistic kinematics
\cite{SCA}, the energy spectrum of a uniformly accelerated particle 
\cite{CGS}, the periodic structure as a function of the momentum p of the neutrino
oscillations \cite{CGS},
the Schwarzschild horizon \cite{TTA},  the expansion
of the very early universe \cite{SPE} and on the value of Higgs--boson mass \cite{KUW}.
It would make the metric observer-dependent, as conjectured by Gibbons
and Hawking \cite{GIB}, and lead in a natural way to hadron confinement \cite{PRE}.

The difficulties of a direct test of Eq. (1.1) obviously reside in the
extreme large value that the normalizing constant $A$ takes
for all known particles
($A\simeq 0.45 \ 10^{30} m$ \ m s$^{-2}$ MeV$^{-1}$). Nevertheless a realistic test that makes use of 
photons in cavities has been recently suggested \cite{INI}. Hopefully
attempts in this direction will lead to some concrete results. In the 
meantime indirect evidence may be gathered by  a variety of different means.

The purpose of this work is to calculate the corrections to the Lamb 
shift of hydrogen or hydrogen-like atoms due to the maximal acceleration.
Because of their relatively simple configuration and of the consequent
precision with which they can be described and observed, these atoms lend 
themselves admirably to the precise test of physical laws. It is therefore 
natural to think in this context of the contribution of maximal 
acceleration, embodied by Eq. (1.1).

The plan of the paper is as follows. Section 2 contains the derivation
of the Dirac Hamiltonian using the tetrad formalism and the
metric (1.1) in a flat background. In Section 3 the Schr\"{o}dinger equation
is obtained from the Dirac equation by splitting the spinor $\psi$ into
large and small components, which is frequently used in calculations
of the Lamb shift. In Section 4 the actual corrections to the levels 
$2s$ and $2p$ are calculated. Section 5
contains the Lamb shift, shift of the states $1s$ and the 
Lamb shift corrections in which the
maximal acceleration is a universal constant. Section 6 contains summary 
and discussion.

\section{The Dirac Hamiltonian}
\setcounter{equation}{0}

If the background metric is flat, Eq. (1.1) becomes simply conformally flat
\begin{equation}
\tilde g_{\mu\nu}=\eta_{\mu\nu}\left(1-\frac{|\ddot{x}|^2}{A^2}\right)=
\sigma^2(x)\eta_{\mu\nu}\,{,}
\end{equation}
where $|\ddot{x}(s)|^2=|\ddot{x}_{\mu}\ddot{x}^{\mu}|$ is the acceleration
field of the particle and $\sigma(x)$ is the conformal factor. The dependence
of $\sigma$ on the variable $x$ implies that the effective geometry defined 
by (2.1) is no longer flat. This has important consequences. In fact one must 
now take into account effects due to the curvature of space-time and
its coupling to the particle itself. 

The electron in the hydrogen atom for the Lamb shift problem is described
by the Dirac equation in flat space-time. With the introduction
of the metric tensor (2.1), the Dirac equation must be generalized to 
curved space-time. This generalization is not trivial, but can be 
accomplished by means of vierbeins that connect a generic non-inertial
frame to a locally Minkowski frame \cite{NAK}. 
Eq. (2.1) above provides immediately
the vierbeins
\begin{equation}
e_{\mu}^{\,\,\,\,a}(x)=\sigma(x)\delta_{\mu}^{\,\,\,\,a}\,{,}
\end{equation}
where Latin indices refer to the local inertial frame and the Greek
indices to the generic non-inertial frame. The covariant Dirac equation is
written in the form
\begin{equation}
(i\hbar\gamma^{\mu}(x){\cal D}_{\mu}-m)\psi(x)=0\,{,}
\end{equation}
where the matrices $\gamma^{\mu}(x)$ satisfy the anticommutation relations
$\{\gamma^{\mu}(x), \gamma^{\nu}(x)\}$
$=2\tilde g^{\mu\nu}(x)$. The covariant
derivative ${\cal D}_{\mu}\equiv \partial_{\mu}+\omega_{\mu}$ 
contains the total 
connection $\omega_{\mu}$ given by
\begin{equation}
\omega_{\mu}=\frac{1}{2}\sigma^{ab}\omega_{\mu ab}\,{,}
\end{equation}
where
\begin{equation}
\sigma^{ab}=\frac{1}{4}\,[\gamma^a,\gamma^b]\,{,}
\end{equation}
\begin{equation}
\omega_{\mu\,\,\,\,b}^{\,\,\,\,a}=(\Gamma_{\mu\nu}^{\lambda}\,
e_{\lambda}^{\,\,\,\,a}-\partial_{\mu}e_{\nu}^{\,\,\,\,a})
e^{\nu}_{\,\,\,\,b}\,{,}
\end{equation}
and
\begin{equation}
\Gamma_{\mu\nu}^{\lambda}=\frac{1}{2}g^{\lambda\alpha}
(g_{\alpha\mu, \nu}+g_{\alpha\nu, \mu}-g_{\mu\nu, \alpha})
\end{equation}
are the usual Christoffel symbols. The usual flat space-time Dirac
matrices are represented by $\gamma^a$. For conformally flat metrics
$\omega_{\mu}$ takes the form
\begin{equation}
\omega_{\mu}=\frac{1}{\sigma}\sigma^{ab}\eta_{a\mu}\sigma_{,b}\,{.}
\end{equation}
By using the transformations
\begin{equation}
\gamma^{\mu}(x)=e^{\mu}_{\,\,\,\,a}\gamma^a\,{,}
\end{equation}
\begin{equation}
\omega_{\mu}=e_{\mu}^{\,\,\,\,a}\omega_a\,{,}
\end{equation}
and the property $e^{\mu}_{\,\,\,\,a}e_{\mu}^{\,\,\,\,b}=\delta_a^{\,\,\,\,b}$,
Eq. (2.3) can be written in the form
\begin{equation}
\left[ i\hbar\gamma^a\partial_a + i\frac{3\hbar}{2\sigma}\gamma^a(\partial_a
\ln\sigma)-m\right]\psi(x)=0\,{.}
\end{equation}
Since in the problem at hand the electron also interacts with the 
electromagnetic field $A_a\equiv (A_0, \vec{A})$, Eq. (2.11) must be
re-written as
\begin{equation}
\left[ i\hbar\gamma^a\left(\partial_a+i\frac{e}{\hbar c}A_a\right)
+i\frac{3\hbar}{2\sigma}\gamma^a(\ln\sigma)_{,a}-m\right]\psi(x)=0\,{.}
\end{equation}
By writing Eq. (2.12) in the form of a Schr\"{o}dinger equation
one arrives at the Hamiltonian
\begin{equation}
H\equiv c\vec{\alpha}\cdot\vec{p}+eA_0+e\vec{\alpha}\cdot\vec{A}+mc^2\beta
-i\frac{3\hbar c}{2}\gamma^0\gamma^a(-\sigma^{-1})_{,a}\,{,}
\end{equation}
where the last term contains all maximal acceleration effects and can be 
written as
\begin{equation}
H^{\prime}\equiv -i\frac{3\hbar c}{2}\gamma^0\gamma^a(-\sigma^{-1})_{,a}=
-i\frac{3\hbar c}{2}
\left(\matrix{\epsilon_0 & \vec{\sigma}\cdot\vec{\epsilon}\cr
\vec{\sigma}\cdot\vec{\epsilon} & \epsilon_0\cr}\right)\,{,}
\end{equation}
where $\epsilon_0\equiv (-\sigma^{-1})_{,0}$ and $\vec{\epsilon}\equiv 
\vec{\nabla}(-\sigma^{-1})$.

The Hamiltonian that can be derived from (2.3) is in general Hermitian
neither with respect to the conserved scalar product of the spinors in 
curved space-time, nor with respect to the flat scalar product \cite{PAR}.
Hermiticity can however be recovered for stationary states of the atom when
the gravitational field changes slowly with time in a locally inertial
rest frame of the atom. 

$H$ given by (2.13) is also not Hermitian 
with respect to the flat scalar product. When one splits the Dirac spinor
into large and small components, the only non-Hermitian
term is however the one proportional to $\epsilon_0$. If $\sigma$
varies slowly in time, or is time-independent, this term may be neglected.

\section{The Maximal Acceleration Corrections}
\setcounter{equation}{0}

The Hamiltonian (2.14) will now be treated perturbatively . The common 
textbook procedure \cite{ITZ} consists in splitting $\psi(x)$ in large
and small components indicated by $\varphi$ and $\chi$, respectively. One
obtains for the perturbation due to $H^{\prime}$
$$
\delta {\cal E}=<nljm|H^{\prime}|nljm>=-i\frac{3\hbar c}{2}\int d^3\vec{r} 
\psi^{\dagger}_{nljm}\left(\matrix{\epsilon_{0} & \vec{\sigma}\cdot
\vec{\epsilon}
\cr \vec{\sigma}\cdot\vec{\epsilon} & \epsilon_0\cr}\right)\psi_{nljm}=$$
\begin{equation}
=-i\frac{3\hbar c}{2}\int d^3\vec{r}[\epsilon_0 (\varphi^{\dagger}\varphi+
\chi^{\dagger}\chi)+\varphi^{\dagger}(\vec{\sigma}\cdot\vec{\epsilon})\chi
+\chi^{\dagger}(\vec{\sigma}\cdot\vec{\epsilon})\varphi]\,{.}
\end{equation}
On introducing the large and small components, related to each other by
\begin{equation}
\chi=-i\frac{\hbar}{c}\frac{\vec{\sigma}\cdot\vec{\nabla}}{2m}\varphi\,{,}
\end{equation}
and integrating by parts, Eq. (3.1) becomes
\begin{equation}
\delta {\cal E}=\delta {\cal E}_0 
-\frac{3\hbar^2}{4m}\int d^3\vec{r}\,\varphi^{\dagger}
[(\vec{\sigma}\cdot\vec{\epsilon})(\vec{\sigma}\cdot\vec{\nabla})+
(\vec{\sigma}\cdot\vec{\nabla})(\vec{\sigma}\cdot\vec{\epsilon})]\varphi\,{,}
\end{equation}
where
\begin{equation}
\delta {\cal E}_0
=-i\frac{3\hbar c}{2}\int d^3\vec{r}\varphi^{\dagger}\epsilon_0
\left[1-\frac{\hbar^2}{4m^2 c^2}(\vec{\sigma}\cdot\vec{\nabla})^2\right]
\varphi\,{.}
\end{equation}
Rearranging the term in square bracket in (3.3) one obtains
\begin{equation}
\delta {\cal E}=\delta {\cal E}_0
-\frac{3\hbar^2}{4m}\int d^3\vec{r}\varphi^{\dagger}
[\vec{\nabla}\cdot\vec{\epsilon}+2\vec{\epsilon}\cdot\vec{\nabla}]\varphi\,{,}
\end{equation}
where for $\vec{\epsilon}\equiv \vec{\nabla}(-\sigma^{-1})$ one must now substitute
\begin{equation}
\vec{\epsilon}=\frac{1}{\sigma^2}\vec{\nabla}\sqrt{1-\frac{|\ddot{x}|^2}{A^2}}=
-\frac{1}{2\sigma^3 A^2}\vec{\nabla}|\ddot{x}|^2\,{.}
\end{equation}
It is now convenient to examine the term $|\ddot{x}|^2$ in detail.
For particles of charge $q$ that move in electromagnetic fields one has
$$|\ddot{x}|^2=\left(\frac{\gamma q}{m}\right)^2
[-(\vec{E}\cdot\vec{\beta})^2
+|\vec{E}|^2+$$
\begin{equation}
+2\vec{E}\cdot(\beta\times\vec{B})+|\vec{\beta}|^2|\vec{B}|^2
-(\vec{\beta}\cdot\vec{B})^2]\,{,}
\end{equation}
where $\gamma=1/\sqrt{1-|\vec{\beta}|^2}$ and $\vec{\beta}=\vec{v}/c$.

In the case of non-relativistic electrons in an electrostatic field, which
is of interest here, Eq. (3.7) reduces to
\begin{equation}
|\ddot{x}|^2=\left(\frac{e}{m}\right)^2|\vec{E}(\vec{r})|^2\,{,}
\end{equation}
while Eq. (3.6) becomes
\begin{equation}
\vec{\epsilon}=-\frac{1}{2\sigma^3}\left(\frac{e}{mA}\right)^2\vec{\nabla}
|\vec{E}(\vec{r})|^2\,{.}
\end{equation}
On neglecting terms of order $A^{-4}$ in the expansion $\sigma\sim 1-
|\ddot{x}|^2/2A^2+\ldots$, and restricting $\vec{E}(\vec{r})$ to central 
electric fields $E(r)=Ze/r^2$, Eq. (3.9) becomes
\begin{equation}
\vec{\epsilon}=2\left(\frac{Ze^2}{mA}\right)^2\frac{\vec{r}}{r^6}\,{.}
\end{equation}

Moreover, $\delta {\cal E}_0=0$ because for electrostatic 
fields $\epsilon_0=0$.
Substituting the result (3.10) in (3.5) one obtains the maximal acceleration
corrections for an electron in an electrostatic field
\begin{equation}
\delta {\cal E}=\delta {\cal E}_1 +\delta {\cal E}_2\,{,}
\end{equation}
where
\begin{equation}
\delta {\cal E}_1=6K\int d^3\vec{r}\, \varphi^{\dagger}\frac{1}{r^6}\varphi\,{,}
\end{equation}
\begin{equation}
\delta {\cal E}_2=-4K\int d^3\vec{r}\, 
\varphi^{\dagger}\frac{1}{r^5}\frac{\partial}
{\partial r}\varphi\,{,}
\end{equation}
and
\begin{equation}
K\equiv \frac{3\hbar^2}{4m}\left(\frac{Ze^2}{mA}\right)^2\,{.}
\end{equation}

\section{The Correction to the Energy Levels $2s$ and $2p$}
\setcounter{equation}{0}

Before calculating the average values of Eqs. (3.12)-(3.13), a few preliminary 
considerations are in order. By virtue of (3.8), the conformal factor
$\sigma(x)$ may be written as
\begin{equation}
\sigma(x)=\sqrt{1-\frac{|\ddot{x}|^2}{A^2}}=
\sqrt{1-\left(\frac{r_0}{r}\right)^4}\,{,}
\end{equation}
where
\begin{equation}
r_0=\sqrt{\frac{Ze^2}{mA}}\sim \sqrt{Z}\,3.3\cdot 10^{-14}m\,{.}
\end{equation}
Eq. (4.1) is real for $r>r_0$. However, it was assumed in the expansion
of the square root (4.1) leading to Eq. (3.10) that $|\ddot{x}|^2/A^2<<1$.
This requires that in the following only those values of $r$ be chosen that 
are above a cut-off $\Lambda$, such that for $r>\Lambda>r_0$ the validity
of the expansion is preserved. The actual value of $\Lambda$ will be 
selected later.

In order to calculate the corrections to the energy levels $2s$ and $2p$,
the explicit expression of the corresponding wave functions are useful.
These are
\begin{equation}
\varphi_{100}(r,\theta,\phi)=R_{10}(r)Y_0^0(\theta,\phi)=\frac{2}{a_0^{
\frac{3}{2}}}
e^{-\frac{r}{a_0}}Y_0^0\,{,}
\end{equation}
\begin{equation}
\varphi_{200}(r,\theta,\phi)=R_{20}(r)Y_0^0(\theta,\phi)=
\frac{2}{(2a_0)^{\frac{3}{2}}}\left(1-\frac{r}{2a_0}\right)e^{-\frac{r}{2a_0}} Y_0^0\,{,}
\end{equation}
\begin{equation}
\left(\matrix{\varphi_{211} \cr \varphi_{210} \cr \varphi_{21-1}\cr}\right)=
R_{21}(r)\left(\matrix{Y_1^1(\theta,\phi) \cr Y_1^0(\theta,\phi) \cr
Y_1^{-1}(\theta,\phi) \cr}\right)=
\frac{1}{\sqrt{3}(2a_0)^{\frac{3}{2}}}\frac{r}{a_0}e^{-\frac{r}{2a_0}}\left(\matrix{Y_1^1\cr
Y_1^0 \cr Y_1^{-1}\cr}\right)\,{,}
\end{equation}
where $Y_l^m(\theta, \phi)$ are the usual spherical harmonics and
$a_0$ is the Bohr radius. The up and down spin wave functions
$$\chi_{\frac{1}{2}}=\left(\matrix{1 \cr 0}\right)\qquad\qquad{\rm and}\qquad\qquad
\chi_{-\frac{1}{2}}=\left(\matrix{0 \cr 1}\right)$$
appear in the average values only in the combinations $\chi_{1/2}^{\dagger}
\chi_{1/2}=1$ and $\chi_{-1/2}^{\dagger}\chi_{-1/2}=1$ and have been omitted
from (4.3) -- (4.5).

From (4.4) and (3.12), (3.13) one finds for $n=2, l=0$
\begin{equation}
\delta {\cal E}_1^{2,0}=
\frac{K}{4a_0^6}\left\{\left[4\left(\frac{a_0}{\Lambda}\right)^3-
8\left(\frac{a_0}{\Lambda}\right)^2+11\left(\frac{a_0}{\Lambda}\right)\right]
e^{-\Lambda/a_0}+11Ei\left(-\frac{\Lambda}{a_0}\right)\right\}\,{,}
\end{equation}
\begin{equation}
\delta {\cal E}_2^{2,0}=\frac{K}{4a_0^6}
\left\{\left[4\left(\frac{a_0}{\Lambda}\right)^2
-10\left(\frac{a_0}{\Lambda}\right)\right]e^{-\Lambda/a_0}-
11 Ei\left(-\frac{\Lambda}{a_0}\right)\right\}\,{,}
\end{equation}
where $K/a_0^6\sim Z^8\,1.03\,\mbox{kHz}$ and
\begin{equation}
Ei(-\mu)=-\int_{1}^{\infty}dx\, \frac{e^{-\mu x}}{x} \, {,} \, \, \, 
\, \, \,\mu > 0\, {.}
\end{equation}
The correction to the level $2s$ is obtained
by summing (4.6) and (4.7)
\begin{equation}
\delta {\cal E}^{2,0}
=\frac{K}{a_0^6}\left[\left(\frac{a_0}{\Lambda}\right)^3
-\left(\frac{a_0}{\Lambda}\right)^2+\frac{1}{4}\left(\frac{a_0}{\Lambda}\right)
\right]e^{-\Lambda/a_0}\,{.}
\end{equation}
Likewise, the explicit expression for the correction to the $2p$ level is:
$$\delta {\cal E}^{2,1}=
   \delta {\cal E}_1^{2,1}+\delta {\cal E}_2^{2,1}=
$$
$$
=6K\int_{\Lambda}^{\infty} dr\frac{1}{r^4}[R_{21}(r)]^2-
4K\int_{\Lambda}^{\infty}drR_{21}(r)\frac{1}{r^3}\frac{\partial}{\partial r}
R_{21}(r)=
$$
\begin{equation}
=\frac{K}{a_0^6}\frac{1}{12}\left(\frac{a_0}{\Lambda}\right)e^{-\Lambda /a_0}
\,{.}
\end{equation}

\section{\bf Lamb Shift, Shift of the States $2p$ and $1s$}
\setcounter{equation}{0}

It now is possible to calculate the contribution to the Lamb shift
$\delta {\cal E}_L=\delta{\cal E}^{2,0}-\delta{\cal E}^{2,1}$. 
The correction is given in our case by

\begin{equation}
\delta {\cal E}_L=\delta{\cal E}^{2,0}-\delta{\cal E}^{2,1}
=\frac{K}{a_0^6}\left[\left(\frac{a_0}{\Lambda}\right)^3-
\left(\frac{a_0}{\Lambda}\right)^2+
\frac{1}{6}\left(\frac{a_0}{\Lambda}\right)\right]
e^{-\Lambda/a_0}\,{.}
\end{equation}
The cut-off $\Lambda$ must now be chosen in a way that is compatible
with the value normally adopted in QED. This is a characteristic length
of the system, the Bohr radius for the hydrogen atom, that cures the
divergences introduced by the radiative corrections. One may tentatively
choose $\Lambda\sim a_0$ in the present calculation. Then
Eq. (5.1) yields
\begin{equation}
\delta {\cal E}_L\sim +Z^8 0.0631\,\mbox{kHz}\,{.}
\end{equation}
For the sake of completeness we also give the maximal acceleration
corrections to the $1s$ ground state Lamb 
shift $\delta {\cal E}^{1,0}=
\delta {\cal E}_1^{(1,0)}+\delta {\cal E}_2^{(1,0)}$. The non-vanishing
contributions are
\begin{equation}
\delta {\cal E}_1^{1,0}=
\frac{8K}{a_0^6}\left\{\left[\left(\frac{a_0}{\Lambda}
\right)^3-\left(\frac{a_0}{\Lambda}\right)^2+2\left(\frac{a_0}{\Lambda}\right)
\right]e^{-2\Lambda/a_0}-4E_1\left(\frac{2\Lambda}{a_0}\right)\right\}\,{,}
\end{equation}
\begin{equation}
\delta {\cal E}_2^{1,0}=\frac{8K}{a_0^6}\left\{\left[\left(
\frac{a_0}{\Lambda}\right)^2-2\left(\frac{a_0}{\Lambda}\right)\right]
e^{-2\Lambda/a_0}+4E_1\left(\frac{2\Lambda}{a_0}\right)\right\}\,{.}
\end{equation}
By subtracting from $\delta {\cal E}^{1,0}$ 
with $\Lambda\sim a_0$ the value
$\delta {\cal E}^{2,0}$ given by (4.9), one obtains the desired shift
\begin{equation}
\delta {\cal E}^{1,0}-\delta {\cal E}^{2,0}\simeq 0.98\,\frac{K}{a_0^6}=
Z^8\, 1.01\, \mbox{kHz}\,{.}
\end{equation}
It has been assumed so far that the maximal acceleration is mass-dependent.
It is also possible to take the view \cite{BRA} that the maximal
acceleration be a universal constant $A_P=m_P c^3/\hbar$, where $m_P=
\sqrt{\hbar c/G}$ is  Planck's mass. In this case $r_0=\sqrt{Ze^2/m A_P}
\sim \sqrt{Z}\, \, 2.13\cdot 10^{-25}m$, and $A_P\sim 5.5\cdot
10^{51}$ $m/s^2$ and
$K/a_0^6\sim Z^8\, 1.28\cdot 10^{-45} \mbox{kHz}$.

Obviously, the Bohr radius can no longer be chosen as cut-off 
in this case because the 
magnitudes of the lengths of physical interest are comparable with Planck's 
length $l_P=\sqrt{\hbar G/c^3}\sim 10^{-35}m$. A better cut-off may be 
derived from the requirement that in the expansion of $\sigma$ terms 
$(r_0/r)^4$ be at least $\sim 10^{-2}$. This leads to $\Lambda\sim 100^{1/4}
r_0\sim 6.76\cdot 10^{-25}$m. The corresponding correction follows from (5.1) 
\begin{equation}
\delta {\cal E}_L\sim Z^8\, 0.61\cdot 10^{-3}\mbox{kHz}\,{.}
\end{equation}

\section{\bf Summary and Discussion} 

The present calculation is based on  the Dirac Hamiltonian 
 (2.13) with the maximal acceleration corrections confined to the 
term (2.14). This has been treated as a perturbation. Legitimate fears about 
the Hermiticity of $H$ have proved unwarranted because of the static nature
of the problem which requires $\epsilon_0 = (-\sigma^{-1})_{,0}=0$.

The standard non-relativistic procedure to split the Dirac wave function
into large and small components has also been followed. This is usually
justifiable because the Lamb shift is a non-relativistic effect at low
frequencies. The results are represented by (5.2), (5.5) and (5.6). 
The latter result pertains to the interpretation, expoused by some 
authors \cite{BRA}, that the maximal acceleration is a universal constant.

A common feature of all results is the dependence on the eighth 
power of $Z$. This would be experimentally significant if the present
uncertainty of about $4\mbox{kHz}$ \cite{ERI} in the measurement of 
$\delta {\cal E}_L$
for $Z=1$ could be extrapolated to atoms of moderately high $Z$, or ionized
atoms. Reported theoretical uncertainties from uncalculated terms for
$Z=1$ amount to $11.0\mbox{kHz}$, which 
is well above (5.2) and could drastically
increase for high $Z$.
The result is nevertheless interesting. Its closeness to (5.6) also indicates
that a Lamb shift measurement would hardly be the way to distinguish 
between the maximal acceleration models of Refs. \cite{ANI} and \cite{BRA}.
The value (5.2) also is smaller than the ground state phase shift, as 
expected on intuitive grounds. Howeover the reported \cite{WEI} measurement
precision of the $1s$ Lamb shift in hydrogen is higher than the $2s-2p$
Lamb shift and at $0.6\mbox{kHz}$ is only a factor $10$ higher than (5.2).
While, on the other hand, the corresponding theoretical 
error is $0.4\mbox{kHz}$,
agreement between the experimental value $8172.874(60)\mbox{MHz}$ and
the theoretical one $8172.802(40)\mbox{MHz}$ still leaves much to be desired.
It is a small consolation that the correction (5.2) is in the right direction.
This also applies to the Lamb shift for the $He^+$ ion \cite{WIJ}. 
In this case (5.2) 
predicts a positive shift of $15.9\mbox{kHz}$ versus 
the present QED deficit of
$\sim 1\mbox{MHz}$. Agreement 
between theory and experiment at higher values of
$Z$ is much worse and so are the corresponding uncertainties.   
The small contribution (5.5) to the fine structure should not be surprising,
given the non-relativistic nature of our approximations.

The values (5.2), (5.5) and (5.6) refer to $\Lambda\sim a_0$ and are rather
conservative. While Lamb shifts and fine structure effects are cut-off 
independent, the values of the corresponding maximal acceleration corrections
increase when $\Lambda$ decreases. This can be understood intuitively because 
the electron finds itself in regions of higher electric field at smaller 
values of $r$. 

A lower cut-off, still compatible with the critical value $r_0$ given by
(4.2), is represented by the Compton wavelength of the electron. This
cut-off, however, yields corrections which are decidedly too high. Both
Lamb shifts become very significant already at values of $\Lambda\sim a_0/2$.
In fact for $Z=1$ one finds $\delta{\cal E}_L\simeq 2.70 \mbox{kHz}$
at $\Lambda=a_0/2$ and $\delta{\cal E}_L\simeq 6.98\mbox{kHz}$ at
$\Lambda=a_0/2.5$. These corrections, added to the theoretical value
$1057849\mbox{kHz}$, yield respectively $1057851.7\mbox{kHz}$
and $1057855.9\mbox{kHz}$, in very good agreement with the experimental 
value $1057851\mbox{kHz}$.
Similary for $Z=1$ the values of $\delta{\cal E}^{1,0}-\delta{\cal E}^{2,0}$
become $20.72\mbox{kHz}$ for $\Lambda=a_0/2$ and $50.95\mbox{kHz}$
for $\Lambda=a_0/2.5$. When added to the theoretical value $8172802\mbox{kHz}$,
they give respectively $8172822.72\mbox{kHz}$ and $8172852.95\mbox{kHz}$,
also in good agreement with the experimental value $8172874(60)\mbox{kHz}$.
The situation is summarized in Table 1.

It is quite remarkable that both $\delta{\cal E}_L$ and 
$\delta{\cal E}^{1,0}-\delta{\cal E}^{2,0}$ have the appropriate signs and
increase with decreasing values of $\Lambda$ as intuitively expected.
In view of the results presented and the approximations used, the cut-off 
$\Lambda\sim a_0/2$ is a better choice.

\begin{center}
\centering{Table 1}\\
All  values reported are in $\mbox{kHz}$
\end{center}

\begin{center}
\begin{tabular}{|c|c|c|c|c|c|}\hline\hline

$\Lambda$ & $a_0$ & $a_0/2$ & $a_0/2.5$ & {\em Theor. value}
           & {\em Exp. value} \\ \hline
$\delta{\cal E}_L$ & $0.0631$ & $2.81$ & $6.90$
     & $1057849(11)$ & $1057851(4)$
\\ \hline
$\delta{\cal E}^{1,0}-\delta{\cal E}^{2,0}$ & $1.01$ & $20.72$
   & $50.95$ & $8172802(40)$ & $8172874(60)$
\\ \hline\hline
\end{tabular}
\end{center}

\bigskip
\begin{centerline}
{\bf Acknowledgments}
\end{centerline}

G.P. gladly acknowledges the constant research support 
of Dr. K. Denford, Dean of Science, University of Regina.

G.L. wishes to thank  Dr K. Denford for his kind hospitality 
during a stay at the University of Regina, where part of this work was
 done.

\bigskip
\begin{centerline}
{\bf REFERENCES}
\end{centerline}

\vfill

\end{document}